\documentclass[aps,prl,twocolumn,superscriptaddress]{revtex4-1}%
\usepackage{graphicx}
\usepackage{amsmath}
\usepackage{amsfonts}
\usepackage{amssymb}
\usepackage{epsfig}
\usepackage{color}
\usepackage{bm}
\usepackage[utf8]{inputenc}
\usepackage{wasysym}
\graphicspath{{pictures/}}
\begin{document}

\title{Synchronized ion acceleration by ultraintense slow light}

\author{A.~V.~Brantov}
\affiliation{P. N. Lebedev Physics Institute, Russian Academy of
Science, Moscow 119991, Russia}
\affiliation{Center for Fundamental and Applied Research,
Dukhov Research Institute of Automatics (VNIIA), Moscow 127055, Russia}
\author{E.~A.~Govras}
\affiliation{P. N. Lebedev Physics Institute, Russian Academy of
Science, Moscow 119991, Russia}
\affiliation{Center for Fundamental and Applied Research,
Dukhov Research Institute of Automatics (VNIIA), Moscow 127055, Russia}
\author{V.~F.~Kovalev}
\affiliation{Center for Fundamental and Applied Research,
Dukhov Research Institute of Automatics (VNIIA), Moscow 127055, Russia}
\affiliation{Keldysh Institute of Applied Mathematics, Russian Academy of Sciences, Moscow 125047, Russia}
\author{V.~Yu.~Bychenkov}
\affiliation{P. N. Lebedev Physics Institute, Russian Academy of
Science, Moscow 119991, Russia}
\affiliation{Center for Fundamental and Applied Research,
Dukhov Research Institute of Automatics (VNIIA), Moscow 127055, Russia}

\begin{abstract}
An effective scheme of synchronized laser-triggered ion acceleration and the
corresponding theoretical model are proposed for a slow light pulse of
relativistic intensity, which penetrates into a near-critical-density
plasma, strongly slows, and then increases its group velocity during
propagation within a target. The 3D PIC simulations confirm this concept for
proton acceleration by a femtosecond petawatt-class laser pulse experiencing
relativistic self-focusing, quantify the characteristics of the generated
protons, and demonstrate a significant increase of their energy compared
with the proton energy generated from optimized ultrathin solid dense foils.
\end{abstract}
\maketitle

With the rapid development of laser technology, several acceleration
concepts using short relativistically intense laser pulses are applied for
generating high-energy ions \cite{Daido_2012_RoPiP_75_056401,Macchi_2013_RoMP_85_020751}.
Most mechanisms of laser-triggered ion generation are tailored to a forward
acceleration of ions from solid targets, but a new trend has recently
appeared in this field based on using low-density targets 
\cite{Passoni_2014_PPaCF_56_045001,Bin_2015_PRL_115_064801},
which could be related to advanced materials such as aerogels, nanoporous
carbon, etc. The hope in using such targets are that they may increase the
energy of the accelerated ions compared with solid targets, particularly
when ions are accelerated by lasers that are now available with $\sim$1~PW
power. One of the challenging ways for accelerating ions up to $\sim$1~GeV
by such lasers is ion wake-field acceleration \cite{Shorokhov_2004_LaPB_22_020175,Shen_2007_PRE_76_055402},
but this is a difficult task for existing high-power laser systems because
heavy particles cannot be pre-accelerated and trapped by the wake field as
easily as electrons can \cite{Esarey_2009_RoMP_81_031229,Pukhov_2002_APB_74_4-50355}.

For an effective acceleration in a rare plasma, similarly to electron
wake-field acceleration, ions must be pre-accelerated up to a velocity
close to the speed of light. Based on simplified 1D and 2D numerical
simulations, several two-stage schemes for ion acceleration have been
proposed \cite{Shen_2009_PRSTAaB_12_121301,Yu_2010_NJoP_12_045021,Zhang_2010_PoP_17_123102,Zheng_2011_EL_95_055005}
to test the idea using an additional target (thin foil or micro-droplets) to
pre-accelerate ions in the radiation-pressure-dominated regime \cite{Esirkepov_2004_PRL_92_175003}.
These ions can then be trapped and accelerated in a gas plasma. For the
proposed scenario to be viable, an Exawatt-class laser is required. The
further development in this direction involved the laser snowplow effect in
a near-critical density plasma, where the electrostatic potential generated
by the laser pulse accelerates and reflects ions \cite{Shorokhov_2004_LaPB_22_020175,Sahai_2013_PRE_88_043105,Wang_2015_PRSTAaB_18_021302}
similarly to collisionless electrostatic shock acceleration \cite{Haberberger_2011_NP_8_010095}.
Specifically, a near-critical-density plasma, which reduces the laser group
velocity, was proposed \cite{Wang_2015_PRSTAaB_18_021302} for a second stage
of accelerating ions initially pre-accelerated from a thin target by a
circularly polarized super-Gaussian pulse. Based on the simulation results,
a proton acceleration to an energy of hundreds of MeV was reported. It is
important that the effect of relativistically induced transparency plays a
key role in ion acceleration with a near-critical density plasma \cite{Sahai_2013_PRE_88_043105}.

A pulse of intense slow light can trap slow or even at-rest ions in its
ponderomotive sheath potential, which acts as a snow plow on electrons and
accelerates ions. But a laser pulse with a group velocity significantly less
than the speed of light cannot accelerate ions to a high energy $\sim$1~GeV
until the group velocity itself starts to increase during pulse propagation.
In this letter, we propose a new scheme for ion acceleration by a
femtosecond laser pulse from a low-density target with an electron density
near the threshold of relativistic transparency. The key point of this
scheme is the capability for the laser pulse first to slow and then to increase
its group velocity monotonically with the propagation distance. This is an
effective way to accelerate ions by a laser ponderomotive field on the
up-going pulse similarly to electron acceleration on the down-going pulse
ramp \cite{Lau_2015_PRSTAaB_18_024401}. The monotonic increase of the pulse
group velocity makes ions achieve a synchronized acceleration by slow light
(SASL). We also present an analytic model and 3D PIC simulations that
demonstrate how SASL works when the increase of the pulse group velocity
during the propagation of a femtosecond petawatt-class laser pulse in a
near-critical-density plasma is due to relativistic self-focusing. Schemes
of something similar to particle--field synchronization have been proposed
by Katsouleas \cite{Katsouleas_1986_PRA_33_032056} (accelerated electrons
and wake field in a rare plasma with a decreasing density) and by Bulanov
et~al.~\cite{Bulanov_2010_PRL_104_135003} (accelerated ions and
electromagnetic field in the radiation-pressure regime).

A relativistic laser pulse can propagate in a plasma with the electron
density $n_e/\gamma\,n_c<1$, where $n_c$ is the electron critical plasma
density and $\gamma$ is the electron gamma factor. The group velocity
$v_g=c\sqrt{1-n_e/\gamma\,n_c}$ ($c$ is the speed of light) of the light
pulse inside the plasma should be small for effective loading and trapping
of the ions in a self-consistent ponderomotive sheath. This is possible only
in a rather narrow range of target densities, $n_e\sim\gamma n_c$, which
slow a laser pulse and allow only the very top of the pulse with a
sufficiently large intensity for relativistic self-transparency to
propagate. For a given laser intensity, we should correspondingly expect a
substantial selectivity of the target density, and this is an expected
specific feature of the acceleration scenario that we discuss.

When the front of a laser pulse hits an overdense target ($n_e>\gamma n_c$),
it penetrates the target to a skin depth $\sim c\,\sqrt{\gamma}/\omega_{pe}$
and pushes electrons by the ponderomotive force $F_p$. This force quickly
moves electrons deeper into the target creating an electron density spike at
the pulse front until $F_p$ is balanced by the charge separation electric
force (see, e.g., \cite{Cattani_2000_PRE_62_011234,Siminos_2012_PRE_86_056404}),
which roughly corresponds to the electron equilibration $eE\sim F_p =-\nabla\Phi$.
The slower the pulse propagation is, the more accurate the estimate $eE=F_p$.
To describe the SASL mechanism qualitatively, we adopt this equality and the 
widely used estimates $\Phi=m_ec^2\gamma$ and $\gamma=\sqrt{1+a^2(x,t)/2}$ for
the ponderomotive potential and gamma factor, where
$a=0.85\,\sqrt{I[10^{18}\,\rm{W/cm}^2]\lambda^2[\rm{\mu m}]}$ is the
normalized amplitude of the laser vector potential. Initially, the
characteristic scale length of the pulse intensity is the skin depth. It
increases in time (as laser pulse intensity increases) and reaches a value
$\sim c/\omega_0$, when the target becomes transparent for the near-peak
intensity. The pulse starts propagating inside the target with a small group
velocity $v_g\ll c$. In some specific cases, the group velocity can increase
as the light penetrates deeper into the plasma, for example, as a result of
relativistic self-focusing or a monotonic decrease of the target density
from front to back. The ponderomotive electric sheath, which propagates in
the plasma with the same group velocity, i.e., $\Phi=m_ec^2\sqrt{1+a^2(x-v_gt)/2}$,
can trap some ions. If the rate of ponderomotive ion acceleration is close
to the rate of the laser pulse acceleration, then the ions gain energy very
efficiently. To justify this, we consider the equations of motion for a test
ion (a proton, for definiteness) accelerated in the traveling ponderomotive
sheath 
\begin{equation}
\frac{dp}{dt}=-\frac{d}{d\xi}\Phi(\xi)\,,\qquad
\frac{dx}{dt}=v=\frac{p}{\sqrt{1+p^2}}\,,
\label{eq_rel}
\end{equation}
where the coordinates $x$ and $\xi(x,t)=x-t\,v_g(t)$ are normalized to the
characteristic pulse spatial width $\sigma$. The time $t$, the velocities
$v$ and $v_g$, the momentum $p$, and the sheath potential $\Phi$ are
respectively normalized to $\sigma/c$, $c$, $m_p\,c$, and $m_e\,c^2/\rho$
($\rho=m_e/m_p =1/1836$).

System of equations~\eqref{eq_rel} describes a nonlinear oscillator and
allows finite and infinite motion. Finite motion corresponds to ion
reflection from the sheath as for acceleration by a collisionless shock
wave \cite{Sagdeev_ShockWaves_VTP}. For example, in the nonrelativistic
limit $p\ll1$, system~\eqref{eq_rel} has a well-known solution in the case
of a constant group velocity $v_g=v_{g,0}$. If the initial proton velocity
$p_0$ is less than $v_{g,0}$, i.e., the parameter $\delta v=p_0-v_{g,0}<0$
and $\Phi_{\max}=\rho\sqrt{1+a_0^2/2}>(\delta v)^2/2$ (cf.~\cite{Wang_2015_PRSTAaB_18_021302}),
then the proton is reflected by the ponderomotive potential and attains the
velocity $2v_{g,0}-p_0$. In the case of an accelerating pulse, i.e., a
time-dependent group velocity $v_g(t)$, the proton after being reflected can
be caught by the pulse and reflected again. Each reflection increases the
proton momentum in accordance with the instantaneous pulse velocity. These
multiple reflections significantly accelerate the ion if it is in phase with
the group velocity increase. We call this the SASL case. The opposite case
of infinite motion corresponds to the absence of any ion reflection from the
potential.

As an example of system~\eqref{eq_rel}, we consider a linearly increasing
group velocity $v_g(t)=v_{g,0}+w_0\,t$ in the nonrelativistic case $p\ll1$.
The corresponding analytic solution $\xi(t)$ is given in quadrature by the
expression
\begin{equation}
t\!-\!t_0\!\!=\pm\!\!\int\limits_{\xi_0}^{\xi}\!\!
\frac{d\,z}{\sqrt{\dot{\xi}_0^2+2(\Phi(\xi_0)-\Phi(z))-4w_0(z-\xi_0)}}\,,
\label{eq_sol_nonrel}
\end{equation}
where $\xi_0=\xi(t_0)$ and $\dot{\xi}_0=\dot{\xi}(t_0)$. Equation~\eqref{eq_sol_nonrel}
in general form describes the ion motion starting at the instant $t_0$ from
the position corresponding to $(\xi_0,\dot{\xi}_0)$. In the case of finite
(oscillatory) motion, there are two return points in the pulse reference
frame where $\dot{\xi}=0$. In the laboratory reference frame, one of the
return points corresponds to the maximum distance of the ion from the sheath
(the point with the maximum coordinate $\xi$), and the second return point
is the electrostatic sheath reflection point. The direction of ion motion
from one return point (or the initial point corresponding to $t_0=0$,
$\xi_0=x_0$, $\dot{\xi}_0=\delta v$) to the other defines the proper sign
choice in Eq.~\eqref{eq_sol_nonrel}.

\begin{figure}[!ht]
\centering{\includegraphics[width=6.8cm]{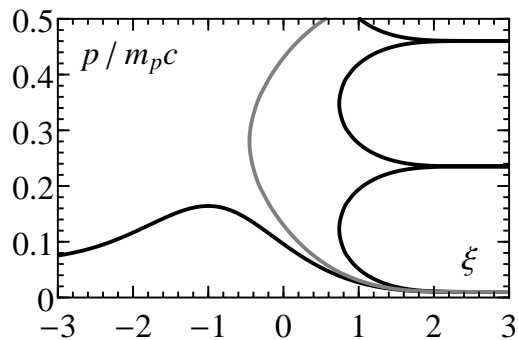}}
\caption{Phase space for an ion starting from $x_0=3$ with $\delta v=0.01$.
The gray curve ($w_0\approx0.028$) separates the finite and infinite regimes
illustrated by the examples $w_0=0.001$ (at the right) and $w_0=0.004$ (at
the left) ($w_0=0.004$). The laser amplitude is $a_0=60$.}
\label{pic_phase}
\end{figure}

\begin{figure} [!ht]
\includegraphics[width=6.8cm]{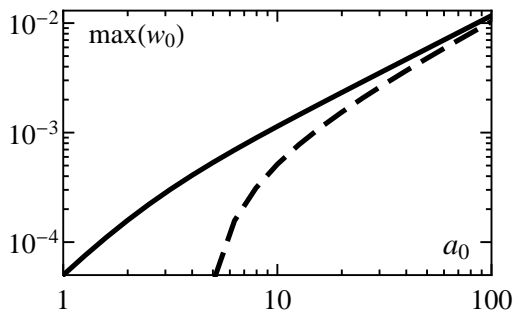}
\caption{Maximum pulse acceleration above which a synchronized ion--pulse
motion is impossible for $\delta v=0$ (solid curve) and $\delta v=\pm0.05$
(dashed curve).}
\label{pic_max_w0_a0}
\end{figure}

To illustrate the obtained analytic solution given by Eq.~\eqref{eq_sol_nonrel},
we present phase plots for the case of a Gaussian spatial profile of the
ponderomotive electric field. Figure~\ref{pic_phase} shows the phase 
portraits for an ion that starts from $x_0=3$ with $\delta v=0.01$ and is
accelerated by an electrostatic sheath with $a_0=60$. The ion momentum is 
constrained by the value $0.5$, when the nonrelativistic solution given by
Eq.~\eqref{eq_sol_nonrel} becomes inapplicable.
Figure~\ref{pic_phase} clearly demonstrates ion oscillations with multiple
reflections from the pulse (SASL) for a sufficiently small pulse
acceleration $w_0$, which allows an effective energy increase. The existence
of two roots of the equation $\dot{\xi}=0$ defines a restriction on the
pulse acceleration $w_0$ for which SASL is allowed:
\begin{equation}
w_0\leq\frac{1}{2(x_0-\xi_*)}
\left[\Phi(\xi_*)-\Phi(x_0)-\frac{(\delta v)^2}{2}\right]\,,
\label{eq_w0limit}
\end{equation}
where the value $\xi_*$ is defined from the solution of the equation
\begin{equation}
\Phi(\xi_*)-\Phi(x_0)=\frac{(\delta v)^2}{2}+(\xi_*-x_0)\Phi'(\xi_*)\,.
\label{eq_xi_ast}
\end{equation}
For a given $w_0$, we can find the domain of initial values $(x_0,\delta v)$
satisfying the SASL condition. Correspondingly, there is a maximum value of
$w_0$ above which a synchronized ion--pulse motion is impossible. The
dependence of the maximum pulse acceleration $\max(w_0)$ on the maximum laser
pulse amplitude is shown in Fig.~\ref{pic_max_w0_a0}.
The steep increase of $\max(w_0)$ with the laser field amplitude for
sufficiently high $|\delta v|$ (i.e., $\delta v=\pm0.05$; see the dashed
curve in Fig.~\ref{pic_max_w0_a0}) in fact demonstrates the existence of the
SASL threshold, i.e., $(\delta v)^2/2=\Phi_{\max}-\Phi(x_0)\,.$ The highest
pulse acceleration is allowed only for protons initially at rest in the
pulse reference frame, $\delta v=0$.

From the first integral of Eq.~\eqref{eq_sol_nonrel},
$\dot{\xi}^2/{2}-{\dot{\xi}_0^{\,2}}/{2}=\Phi(\xi_0)-\Phi(\xi)-2\,w_0(\xi-\xi_0)\,$,
we can estimate the proton energy increase. For protons initially at rest,
the proton energy evolves as $\varepsilon=2w_0(vt-x)+\Phi(\xi_0)-\Phi(\xi)\,$.
In average, an ion moves together with the laser pulse and has a constant
acceleration $x\sim w_0t^2$, which results in an ion energy increase as
$w_0^2t^2$. We note this time dependence coincides with that for a Coulomb
explosion of a plane target, which yields the maximum possible energy gain.
But such a rapid energy gain is only in the nonrelativistic limit.

\begin{figure} [!ht]
\centering{\includegraphics[width=6.8cm]{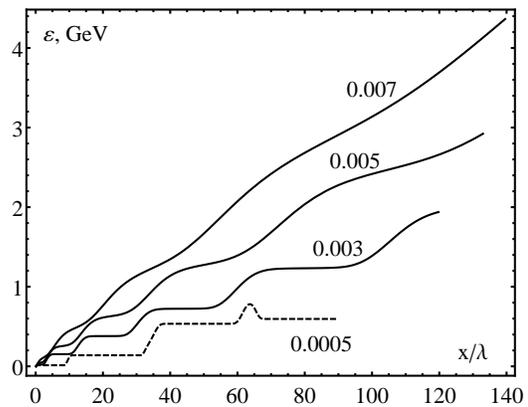}\hspace{0.6cm}}
\caption{The proton energy gain for $x_0=1$ and $p_0=0$ versus the
propagation length for the ponderomotive potential characterized by $a_0=60$
and $v_g=w_0t/(1+w_0t)$, where $w_0=0.003,0.005,0.007$ (the numbers near the
curves). The dashed curve shows the energy gain for $a_0=60$ and $v_g=0.0005t$.}
\label{figm1}
\end{figure}
The nonrelativistic solution of system~\eqref{eq_rel} can describe ion
acceleration only for $\varepsilon\ll m_pc^2$, while the SASL can in
principle yield a much higher ion energy. But for a linear time dependence
of $v_g$, the relativistic nonlinearity in system~\eqref{eq_rel} sooner or
later breaks the SASL. The test ion acceleration decreases, and the
ponderomotive potential of the laser pulse overcomes the ions. This occurs
after three consecutive reflections from the sheath and is illustrated by
the bump on the dashed line in Fig.~\ref{figm1}. To describe the
relativistic regime more accurately, we must take the natural saturation of
the group velocity at the speed of light into account. We present the
corresponding illustration in Fig.~\ref{figm1} as a result of solving
system~\eqref{eq_rel} numerically for the saturating group velocity
$v_g=w_0t/(1+w_0t)$, which shows continuous particle acceleration up to
relativistic energies. The average energy of accelerated protons increases
almost linearly with the propagation distance. Certainly, such a theoretical
illustration cannot claim to be a proof of the new approach, but it at least
vividly highlights the possible SASL advantage.

Other physical effects accompanying the laser pulse propagation, for
example, pulse depletion, deformation of the pulse shape, violation of the
ideal electron equilibration $eE\sim F_p$, complicate the theoretical model.
They might degrade the theoretical estimates of the ion energy gain, but the
basic principle of the SASL should remain. To confirm this, we performed 3D
PIC simulations of proton acceleration by short (the FWHM pulse duration is
30~fs), tightly focused (the FWHM transverse size is $4\lambda$) Gaussian
laser pulses using the code MANDOR \cite{Romanov_2004_PRL_93_215004}. The
laser intensities in the focal spot were $I=5\times10^{20}$~W/cm$^2$ and
$I=5\times10^{21}$~W/cm$^2$, which correspond to the respective laser
energies 3~J ($a_0=19.1$) and 30~J ($a_0=60$) for $\lambda=1~\mu$m. The
laser pulse was focused on the front side of a thin CH$_2$ plasma target,
which consists of electrons, hydrogen ions, and fully ionized carbon ions
(C$^{6+}$). The target densities were decreased from the solid mass density
of CH$_2$ ($1.1$~g/cm$^3$), which corresponds to $n_e=200n_{c}$ to the
density $5.5$~mg/cm$^3$ ($n_e=n_{c}$). The target thickness $l$ was varied
from $3$~nm to $10$~$\mu$m.

\begin{figure} [!ht]
\centering{\includegraphics[width=6.8cm]{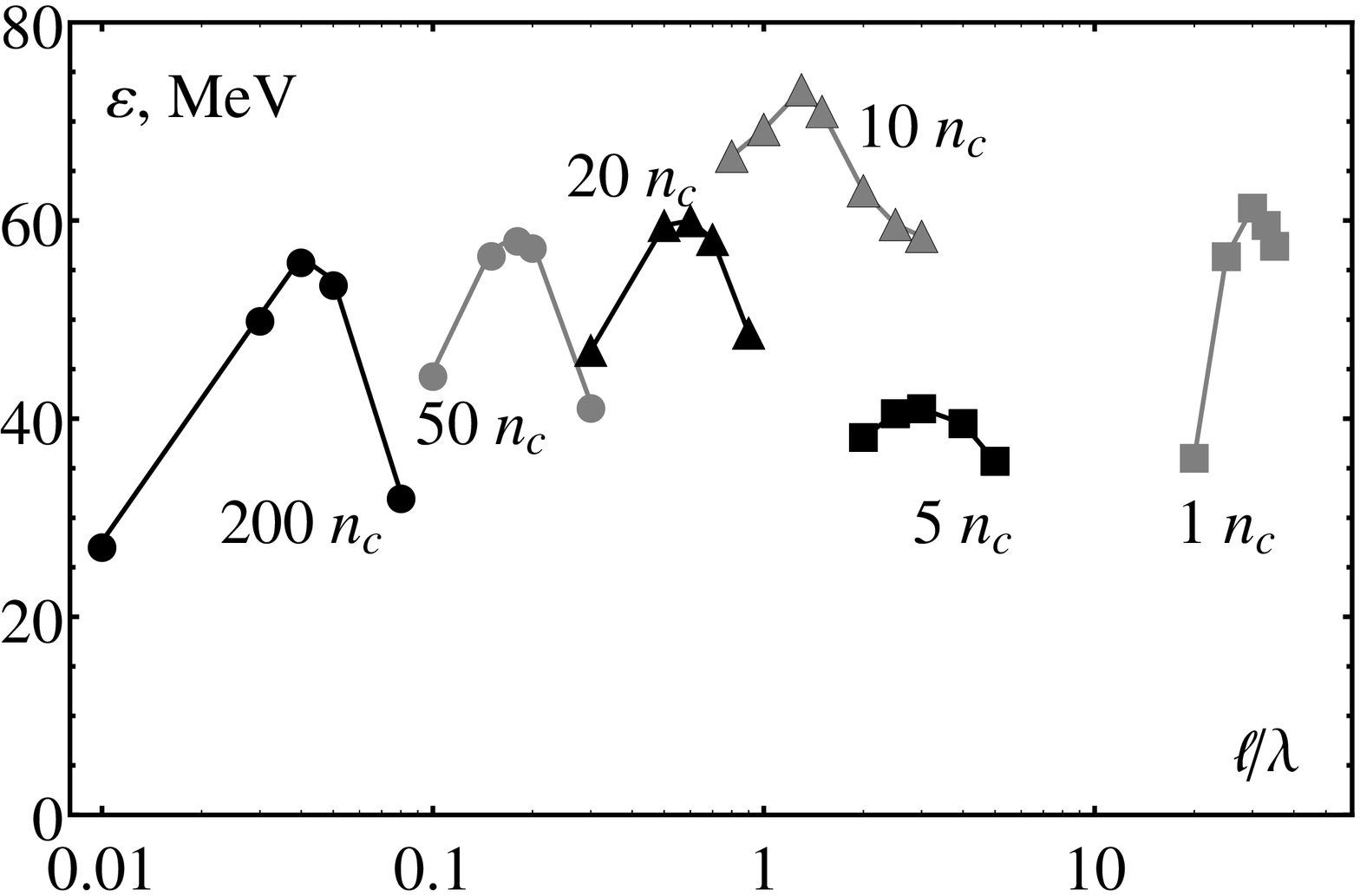}
\hspace{0.6cm} \includegraphics[width=7.cm]{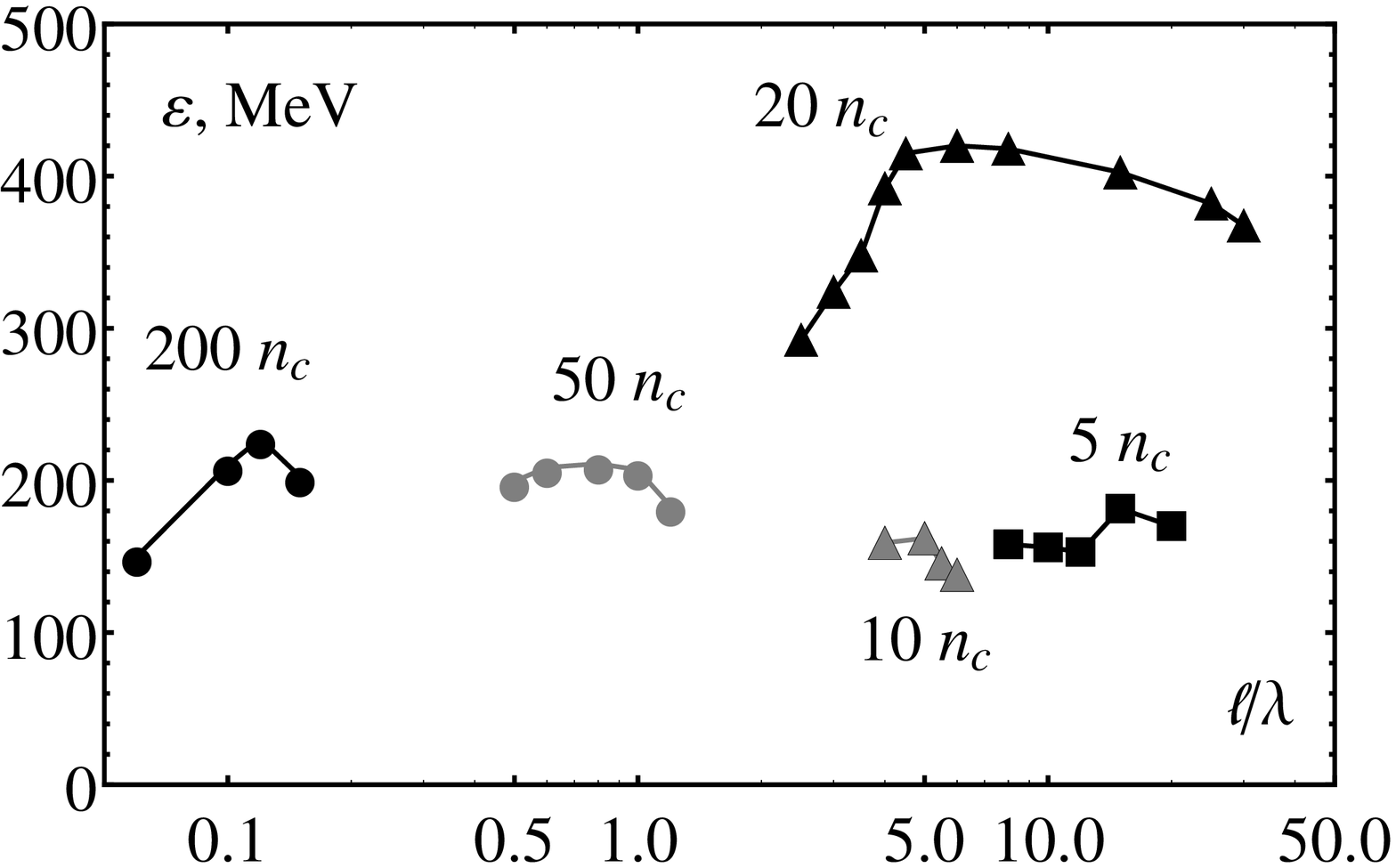}}
\caption{Maximum proton energy versus target thickness for an electron
target density of $200n_c$ (black points), $50n_c$ (gray points), $20n_c$
(black triangles), $10n_c$ (gray triangles), $5n_c$ (black squares), and
$n_c$ (gray squares) and for laser energies of 3~J (top panel) and 30~J
(bottom panel).}
\label{fig1s}
\end{figure}

\begin{figure}[!ht]
\centering{\includegraphics[width=6.8cm]{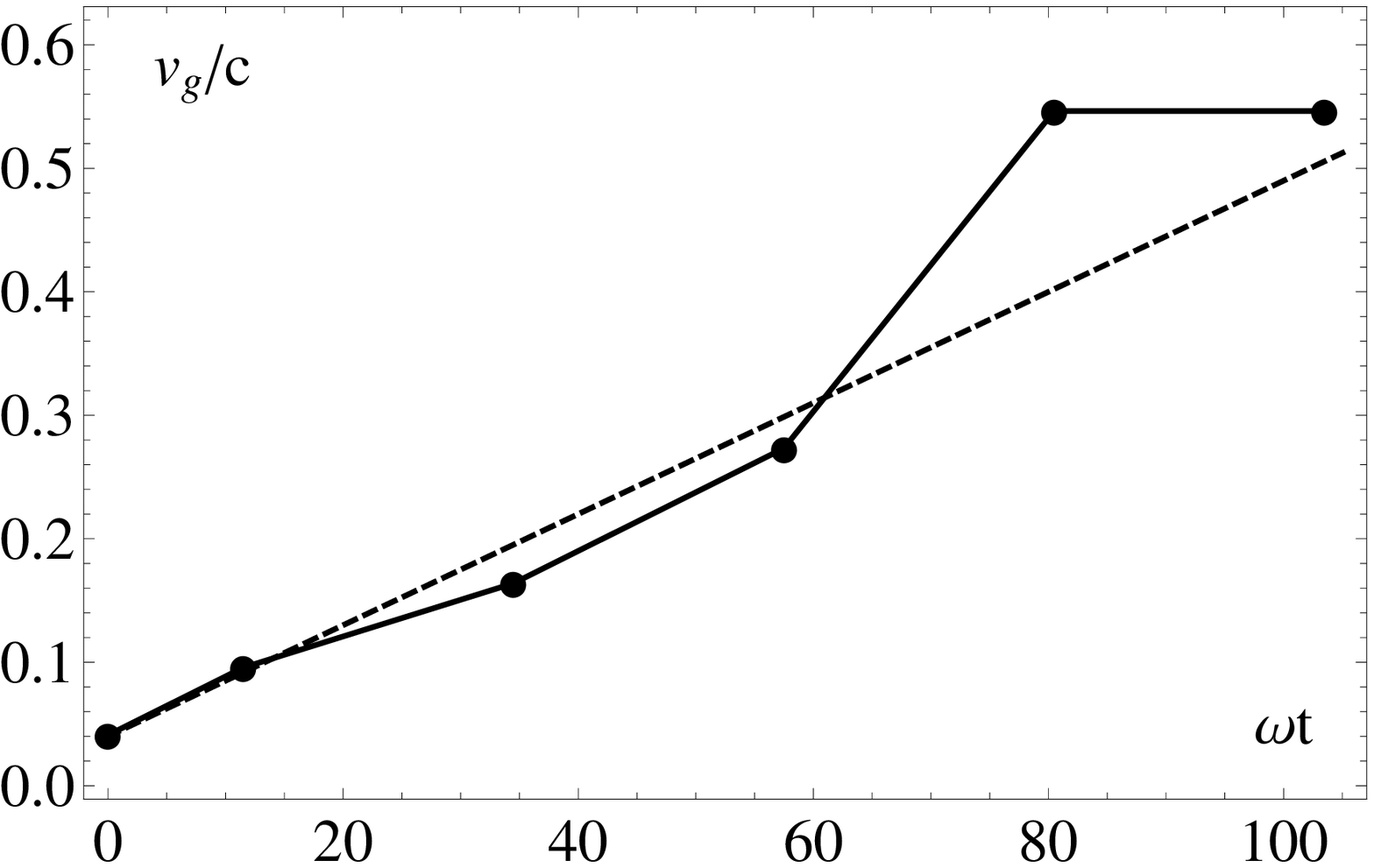}
\hspace{0.6cm} \includegraphics[width=7. cm]{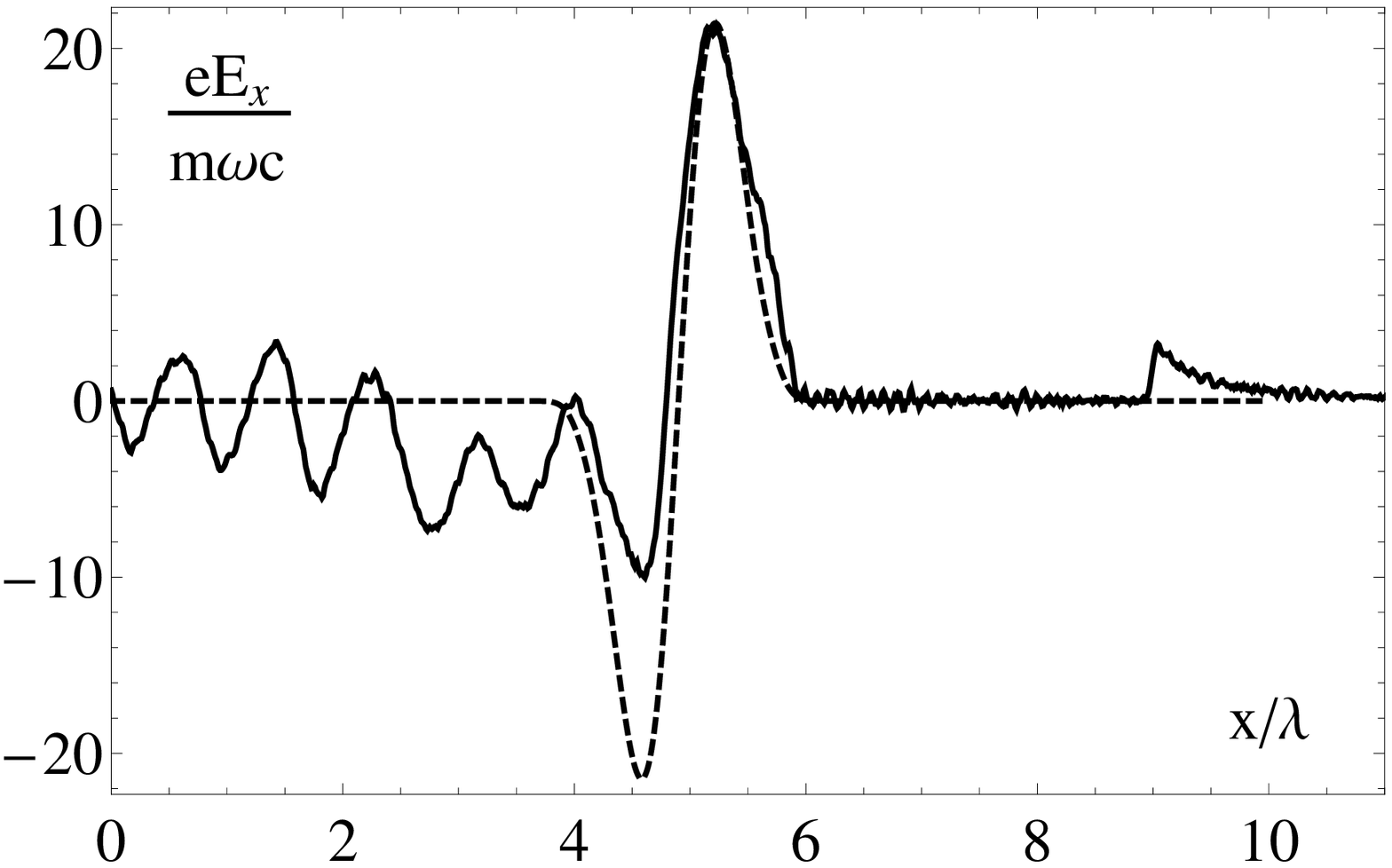}}
\caption{Pulse velocity inside a $6~\mu$m target with an electron density
$20n_c$ for a laser pulse with an energy of 30~J (top panel) and an
electrostatic field (bottom panel). The target is placed from $x=3\lambda$
to $x=9\lambda$. The dashed curve corresponds to the Gaussian fit of the
potential.}
\label{fig2s}
\end{figure}

\begin{figure} [!ht]
\centering{\includegraphics[width=6.8cm]{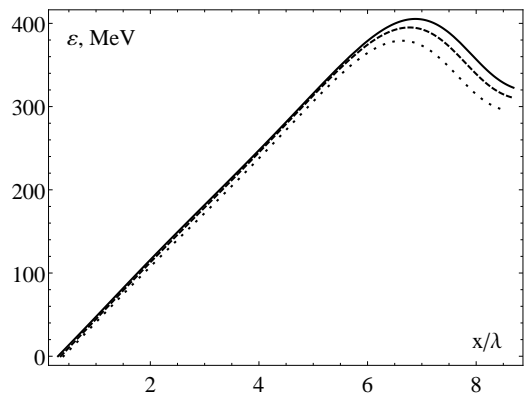}}
\caption{Energy gain of test protons with $x_0=0.4\lambda$ and $p_0=0$
(dotted curve), $x_0=0.3\lambda$ and $p_0=0.01$ ($E_0= 50$~keV) (solid
curve), and $x_0=0.35\lambda$ and $p_0=0$ (dashed curve).}
\label{fig3s}
\end{figure}

Several runs with different target thicknesses were performed for each
target density. The simulation results are presented in Fig.~\ref{fig1s},
which confirms the existence of an optimal target thickness in each case.
Using a laser pulse with an energy of 3~J does not effectively synchronize
the proton acceleration and laser pulse group velocity increase for any
density, although there is a small proton energy increase (up to 30\%).
Proton acceleration by a petawatt-class laser with an energy of 30~J is more
effective. For a target with an optimal density of $20n_c$, a significant
increase of the maximum proton energy was found. For such a density, the
front wing of the laser pulse does not penetrate into the target, but the
near-peak intensity does penetrate and propagates inside the plasma with an
increasing group velocity (the solid curve in the top panel of
Fig.~\ref{fig2s}), which can be roughly approximated as $v_g=c(0.04+0.0045\omega t)$
(the dashed line in the top panel of Fig.~\ref{fig2s}). The laser-induced
electrostatic sheath field corresponds well to the Gaussian-shape potential
with $a_0=100$ and $\sigma=2\omega/c$ (the dashed curve in the bottom panel
of Fig.~\ref{fig2s}). For these parameter values, the increase of the pulse
group velocity is sufficiently small ($w_0=0.009$), which is favorable for
an effective trapping of the protons in the ponderomotive potential
(cf.~Fig.~\ref{pic_max_w0_a0}).
The protons gain significant energy as a result of the SASL mechanism. We
estimated the energy gain of the test protons (initially at rest or weakly
pre-accelerated) in the electrostatic field shown by the dashed curve in
Fig.~\ref{fig2s} as the solution of system~\eqref{eq_rel}. At the distance of
$6~\mu$m, the protons gain energy of $350$ to $400$~MeV (see
Fig.~\ref{fig3s}), which is consistent with the PIC simulation result. For
$x>6 \lambda$, the electrostatic potential overtakes the protons and 
decelerates them as a result of the transition to the relativistic regime
and violation of the phasing-in.

In conclusion, we have proposed a new model of proton acceleration by an
ultraintense slow light pulse interacting with low-density targets. The key
points of this mechanism are to stop the laser pulse at the front of the
target and then accelerate the infiltrating intense part of the pulse inside
a plasma at the same rate as the proton energy increase in a ponderomotive
potential to achieve synchronized acceleration by slow light (SASL). In the
case considered, the linearly polarized laser pulse propagates and increases
its group velocity as a result of a relativistic self-induced transparency.
Another scheme of pulse acceleration could be based on using a plane target
with its density decreasing from front to back. The SASL regime is
challenging for low-density material applications and will require
production of lightweight foils with fully variable and controllable
parameters. Such foils should typically have a multimicron thickness that
makes them more robust in laser acceleration experiments in contrast to
nanoscale solid dense foils, which require an extremely high intensity
contrast ratio. Finally, we note that our simulations with low-density
targets have demonstrated more than a twofold increase of proton energy
compared with the increase in the case of solid dense foils of optimal
submicron thickness \cite{Brantov_2015_PRSTAaB_18_021301}.

This research was funded by a grant from the Russian Science Foundation
(Project No.~14-12-00194).




\end{document}